\documentstyle[seceq,preprint]{jpsj}
\def\runtitle{ Origin of Magic Angular Momentum in a Quantum Dot }
\def\runauthor{Taku {\sc Seki}, Yoshio {\sc Kuramoto}, and Tomotoshi {\sc Nishino}}
\title
{Origin of Magic Angular Momentum in a Quantum Dot under Strong Magnetic Field
} 
\author
{
Taku {\sc Seki},\footnote{present address: Information Systems Division, Hitachi Ltd., 890 Kashimada, Kawasaki 211} Yoshio {\sc Kuramoto}\footnote{e-mail address: kuramoto@cmpt01.phys.tohoku.ac.jp} and Tomotoshi {\sc Nishino}
}
\inst
{
Department of Physics, Tohoku University, Sendai 980-77
}
\recdate{\hspace{3cm}
}
\abst
{
This paper investigates origin of the extra stability associated with particular values (magic numbers) of the total angular momentum of electrons in a quantum dot under strong magnetic field.
The ground-state energy, distribution functions of density and angular momentum, and pair correlation function are calculated in the strong field limit by numerical diagonalization of the system containing up to seven electrons.
It is shown that the composite fermion picture explains the small magic numbers well, while a simple geometrical picture does better as the magic number increases.
Combination of these two pictures leads to identification of all the magic numbers.  Relation of the magic-number states to the Wigner crystal and the fractional quantum Hall state is discussed.  
}
\kword
{
quantum dot, fractional quantum Hall state, composite fermion, exact diagonalization, magic number, Wigner crystal
}

\begin{document}
\sloppy
\maketitle
%

\section{Introduction}

The two-dimensional electron system with a small number $N$ of electrons is realized as the quantum dot formed at a semiconductor interface.\cite{ashoo}  Correspondingly there is a growing theoretical interest in the system.
It has been found \cite{laughlin83,girvin,trugman,maksym,rejaei,maksym2,eric,jain6,lai} that if a system is placed in a strong magnetic field, extra stability arises for a special set of angular momentum $M.$
These values of $M$ are called magic numbers. 
According to numerical diagonalization, the magic number states occur with an interval $\Delta M =N$ for systems with $N \leq 5$, and that these states have a polygonal pattern with $N$ apexes in the configuration of electrons.\cite{maksym,maksym2}
The magic number states prevail the ground-state phase diagram in the plane of magnetic field vs the confining potential for $N=5$ and 6.\cite{eric}
On the other hand, it has been shown\cite{rejaei,jain6} that many of the magic number states can be explained in terms of the composite fermion (CF) picture\cite{jain4} for the fractional quantum Hall (FQH) states. 
The overlap is close to unity  \cite{jain6,jain5} between a trial wave function based on the CF theory and the exact ground state wave function in the presence of the Coulomb interaction.
However, some magic number states do not fit into the interpretation of the CF theory.

The purpose of this paper is to clarify the origin of the magic numbers on the basis of exactly derived results for the energy, and one- and two-body distribution functions. 
In addition to the previously known series with $\Delta M=N$, we find another series with $\Delta M=N-1$  of the magic number states for $N\geq 6$.  We provide simple geometrical interpretation on the origin of new series of magic numbers.  Combining with previous interpretations, we come to  unified understanding of all the magic numbers.

The paper is organized as follows: In \S 2 we define the model which involves truncation to the lowest Landau level.  Section 3 presents numerical results on the ground state energy as a function of the total angular momentum.  The magic numbers are identified for systems with 6 and 7 electrons.  The results on one- and two-body distribution functions  are shown in \S 4 where the characteristic features of the magic-number states are made explicit.  
We provide in \S 5 a geometrical interpretation of the magic number states, and discuss the relation to the CF picture.\cite{jain6}  In \S 6 we combine the geometrical and CF interpretations, and point out the range of validity for each interpretation.  This reveals significance of magic-number states as showing crossover from an incipient fractional quantum Hall state to an incipient Wigner crystal.

\section{Model and Choice of Basis}

We consider a model for a quantum dot in two dimensions with the Hamiltonian 
\begin{equation}
H=H_{0}+V_{1}+V_{2}.
\end{equation}
Here $H_{0}$, $V_{1}$, and $V_{2}$ are given by
\begin{equation}
H_{0}=\frac{1}{2m^{*}}\sum_{i=1}^{N}(\frac{\hbar}{i}\nabla_{i}+\frac{e}{c}{\bf A}_{i})^{2},
\end{equation}
\begin{equation}
V_{1}=\frac{1}{2}\,m^{*}\omega^{2}\sum_{i=1}^{N}|z_{i}|^{2}, \hspace{0.5cm} 
V_{2}=\frac{e^{2}}{\varepsilon_{0}}\sum_{i<j}^{N}\frac{1}{|z_{i}-z_{j}|},
\end{equation}
with $\omega$ being the frequency of the harmonic oscillator, and $z = x+iy$.  Other notations are the standard ones.
The vector potential is chosen to be
$ {\bf A}_i = (y_iB/2, -x_i B/2)$.
We assume that the magnetic field $B$ is so strong that all electrons are in the lowest Landau level (LLL), with spins being completely polarized.
One-electron state is specified by its orbital angular momentum only.  We ignore the spin degrees of freedom and mixing between the LLL and higher Landau levels. 
Hence the Zeeman term has been dropped in our model.

The normalized single electron wave function in the LLL is given by
\begin{equation}
\phi_{m}(z)=\frac{1}{\sqrt{2\pi \ell^{2}2^{m} m!}}\,z^{m}\exp(-\frac{|z|^2}{4\ell^{2}}),
\end{equation}
where $m$ is the angular momentum and $\ell$ the magnetic length defined by
$
\ell=(\hbar c/eB)^{\frac{1}{2}}.
$
The Hamiltonian in the second quantization is written as
\begin{eqnarray}
H&=&\sum_{m=0}^{\infty}<m|H_{0}|m>a_{m}^{\dagger}a_{m}+\sum_{m=0}^{\infty}<m|V_{1}|m>a_{m}^{\dagger}a_{m} \nonumber \\     
&+&\frac{1}{2}\sum_{m_i}<m_{1}m_{2}|V_{2}|m_{3}m_{4}>a_{m_{1}}^{\dagger}a_{m_{2}}^{\dagger}a_{m_{3}}a_{m_{4}},
\end{eqnarray}
where $a_{m}$ is the annihilation operator of an electron with $m$.  The matrix elements are given by
\begin{equation}
<m|H_{0}|m>=\frac{1}{2}\hbar\omega_{c},
\end{equation}
\begin{equation}
<m|V_{1}|m>=\frac{1}{2}m^{*}\omega^{2}\ell^{2}(m+1),
\label{V1}
\end{equation}
and the Coulomb matrix element is computed numerically from
\begin{full}
\begin{eqnarray*}
 <m_{1}m_{2}|V_{2}|m_{3}m_{4}> = (\frac{e^{2}}{\varepsilon_{0}\ell})\,\frac{2^{-d}}{\sqrt{(j-k)!k!(k+d)!(j-k-d)!}} \frac{\Gamma(j-k+1)\,\Gamma(k+d+1)}{\Gamma(d+1)^{2}} \\
\times \int_{0}^{\infty}{\rm d}x\,x^{2d} 
_{1}F_{1}(j-k+1;d+1;-\frac{x^{2}}{2}) _{1}F_{1}(k+d+1;d+1;-\frac{x^{2}}{2}).
\end{eqnarray*}
\end{full}
\noindent Here $_{1}F_{1}$ is Kummer's hypergeometric function, and $m_{1}=j-k,m_{2}=k, m_{3}=k+d, m_{4}=j-k-d$ with $d \geq 0$. 
We discard in the following the constant kinetic term $<m|H_{0}|m>=\frac{1}{2}\hbar\omega_{c}$.
The confinement parameter $\gamma$ is defined by 
\begin{equation}
\gamma=\frac{m^{*}\omega^{2}\ell^{2}}{2e^{2}/(\varepsilon_{0}\ell)}.
\end{equation}
The length is scaled by $\ell$, and energy is scaled by $e^{2}/(\varepsilon_{0}\ell)$ from now on.  

Since the confining potential and the Coulomb interaction have the rotational symmetry, the total angular momentum $M$ can be used to label the many body eigenstates. 
We introduce the distribution function of angular momentum by
\begin{equation}
n_{m}=<\Phi_{0}|a_{m}^{\dagger}a_{m}|\Phi_{0}>,
\end{equation}
where $|\Phi_{0}>$
is the normalized ground state.  Then the total angular momentum is given by
\begin{equation}
M=\sum_{m=0}^{\infty}m n_m.
\end{equation}
Next we introduce a field operator $\psi(z)$ truncated within the LLL:
\begin{equation}
\psi(z)=\sum_{m}\phi_{m}(z)a_{m}.
\end{equation}
Then the charge density $n(z)$ is given by
\begin{equation}
n(z)=<\Phi_{0}|\psi^{\dagger}(z)\psi(z)|\Phi_{0}>, 
\end{equation}
 and the total number $N$ by its integral over the whole space.
Finally the pair (two-body) correlation function is defined as
\begin{equation}
n(z,z')=<\Phi_{0}|\psi^{\dagger}(z)\psi^{\dagger}(z')\psi(z')\psi(z)|\Phi_{0}>.
\end{equation}

\section{Two Series of Magic Numbers}

By using the Lanczos method to diagonalize the Hamiltonian numerically, we calculate the lowest energy and corresponding eigenvectors for each $M$  and $N$.  For given $N$ and $M$, the one body angular momentum takes a value from 0 to $M-N(N-1)/2$.  The upper limit less than $M$ comes from the Pauli exclusion principle.
In constructing the Hamiltonian matrix, we take all the necessary basis for $m$ without truncation. 
We note the following relation:
\begin{equation}
V_1|\Phi_{0}> = \gamma (M+N)|\Phi_{0}>,
\label{V1t}
\end{equation}
which follows from eqs.(\ref{V1}).
Thus the wave function $|\Phi_{0}>$ is actually independent of $\gamma$, and the numerical result for $\gamma =0$ is sufficient to derive the energy for other cases of $\gamma$.

Figure 1(a) shows the results for the ground-state energy vs $M$ in the case of $N=6$.
\begin{figure}
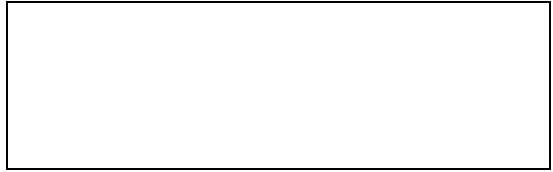

\figureheight{2cm}
\caption{The ground-state energy vs the total angular momentum in the case of (a) $N=6$ and (b) $N=7$.  The parameter $\gamma$ represents the strength of the harmonic confinement as explained in the text. }
\end{figure}
A number of downward cusps appear clearly.  Upon closer inspection one finds that a part of the cusps with  $M= 15, 21, 27, 33, 39, 45$ are represented by the formula
\begin{equation}
M = \frac12 N(N-1)+N k, \ \ (k=0,1,\cdots) 
\end{equation}
with $N=6$. 
This set of states with the interval $\Delta M = N$ are called the series $S_N$ in this paper.  
The other set of cusps appear for states with $M = 25, 30, 35, 40, 45, 50$ with another interval $\Delta M = N-1$ which is called the series $\tilde{S}_N.$  The particular state with $M=45$ belongs to both  $S_N$ and $\tilde{S}_N.$ 
According to Fig.1(a), $M=20$ in the series $\tilde{S}_N$ is not actually a magic number.
With $\gamma =0$, each magic-number state and the adjacent state with one more angular momentum often have the same energy.  This degeneracy comes from the degrees of freedom associated with the center of mass motion as pointed out in ref.\citen{trugman}.

The angular momentum of the absolute ground state is determined by competition between the Coulomb repulsion and the strength of the confinement .  As $\gamma$ increases, the optimum angular momentum shifts to smaller values.  In the case of $\gamma = 0.04$ for example, the state with $M=45$ takes the absolute minimum of the energy.

In Fig.1(b) we show the ground-state energy in the case of $N=7$.  We find clear downward cusps at the series $\tilde{S}_N$.  This includes $M=21,33,39,45,51,57,63$ and 69.  However, the number $M=27$ belonging to $\tilde{S}_N$ is not a magic number. 
On the other hand, the cusps corresponding to the series $S_N$ are not clear except for $M=28$.  This is in strong contrast with the case of $N\leq 6$.
We note that a magic number state with $M=36$ is seen in Fig.1(b).  This state belongs to neither the series $S_N$  nor $\tilde{S}_N$ .  As we discuss later, the CF picture can interpret the magic number $M=36$.

\section{Distribution Functions}
\subsection{Angular momentum}

In order to clarify the electronic property associated with the magic-number states, we calculate the distribution function $n_m$ of the angular momentum.
In this section, we report mainly on results for $N=6$.
Figure 2 shows representative results.
\begin{figure}
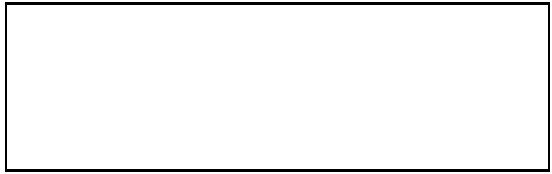

\figureheight{2cm}
\caption{One-body angular momentum distribution function for (a) $M=33$ and 35 and (b) $M=39$ and 45.   The distribution with double peaks as for $M=35$ and 45 is called the type-D, while the one with a single peak as for $M=33$ and 39 is called the type-S.}
\end{figure}
As $M$ increases, two types of distribution appear by turns in the magic-number states.
One is the distribution with double peaks at the origin as well as at a finite angular momentum.  We call the double-peaked distribution the type-D  hereafter. This is the case with $M=35$ and 45. 
The other type is the distribution with a single peak which we call the type-S, as for $M=33$ and 39.
An important observation is that the magic number states with the type-D distribution are all in the series $\tilde{S}_N$, while those with the type-S distribution are all in the series $S_N$.  
These features are common to cases other than $N=6$.

 It is instructive to interpret the result in terms of the Laughlin wave function $\Psi_L (z_1,\ldots ,z_N)$ for finite $N$.  It is given for general $N$ by
\begin{equation}
\Psi _L (z_1,\ldots ,z_N)= \prod _{i<j} (z_i-z_j)^p \exp(-\frac14\sum_{k=1}^N |z_k|^2).
\end{equation}
The odd integer $p$ is related to the filling $\nu = 1/p$ of the LLL in the limit of $N\rightarrow\infty$. 
In the case of finite $N$, the maximum one-body angular momentum $m_{\rm max}$ is given by
$ m_{\rm max} =  pN(N-1)/2 $.
With $p =3$ and $N=6$, in particular, we obtain $m_{{\rm max}}=15$ and the total angular momentum $M$ is given by $M=3N(N-1)/2 =45$.
We note that the numerically obtained results for $M=45$ in Fig.2(b) has the maximum $15$ of the significantly occupied angular momentum in good correspondence to the Laughlin wave function with $p=3$.

\subsection{Charge density}

The larger angular momentum corresponds to the wave function  more extended from the origin.
More explicitly we have the relation
\begin{equation}
<m|r^{2}|m>=2\ell^{2}(m+1).
\label{r2}
\end{equation}
We can therefore expect that $n(z)$ of the type-D distribution should have double peaks both near the origin and near the edge of the dot, and that of the type-S distribution should have only a single peak near the edge. 
We have actually calculated the charge densities for these states with $N=6$ and 7.
Since the result $n(z)$ depends only on $r=|z|$, the density is written as $n(r)$ in the following.  

Figure 3 shows some exemplary results.
\begin{figure}
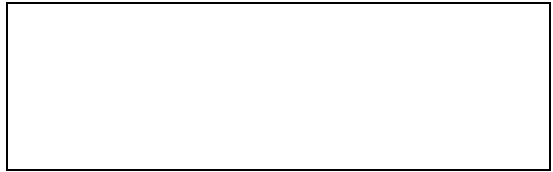

\figureheight{2cm}
\caption{The charge density $n(r)$ vs the distance $r$ from the center of the dot in the case of $M=33$, 35 and 45. }
\end{figure}
It is found that the state with the type-S distribution ($M=33$) indeed has a single peak near the edge of the quantum dot, and those with the type-D distribution ($M=35, 45$) have double peaks.
This feature is common to systems with different electron numbers. 
Since the Laughlin-type state with $M = pN(N-1)/2$  belongs simultaneously to series $S_N$ and $\tilde{S}_N$, it is natural to expect that the charge distribution is also a superposition of the type-D and the type-S distributions. As a result, the distribution should have double peaks.
The result in Fig.3 shows that this is indeed the case with $M=45$ and $N=6$.

It is seen from eq.(\ref{r2}) that the average density of electrons in the quantum dot decreases as the total angular mometum $M$ increases.  In \S 6 we use this fact for unified understanding of magic numbers.

\subsection{Pair correlation function}

In order to see the correlated motion of electrons more closely
we calculate the pair correlation function $n(z,z')$ exactly.  This quantity gives us a quantum analogue of a snapshot picture of electrons.  More precisely, it gives the distribution of electrons on condition that one of the electrons is nailed down at $z'$.
The pair correlation function for $N=5$ or less has been calculated by Maksym,\cite{maksym2} who found a pattern corresponding to the polygon with $N$ apexes for the series $S_N$.
For $N=6$ or more, however, we are not aware of exact results reported so far.  Moreover nothing is known about the pair distribution for the series $\tilde{S}_N$.
We define $r_0$ as the peak position of the charge density, which in the case of type-D distribution is to be taken at the position of the outer peak.
Then we calculate $n(z,z')$ numerically, setting $|z'| = r_0$.

From Fig.4 with $N=6$, it is clear that electrons are arranged like a hexagon in the magic number state ($M=33$) belonging to the series $S_N$, and like a pentagon in the magic number state ($M=35$) of the series $\tilde{S}_N$.  
\begin{figure}
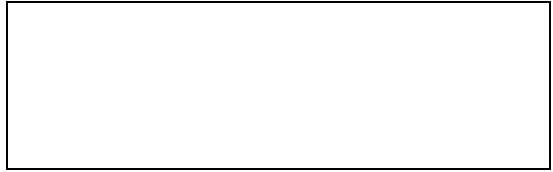

\figureheight{2cm}
\caption{Contour map of the pair correlation function $n(z,z')$ for (a) $M=33$ and (b) $M=35$.  The complex coordinate $z'=r_0 i$ corresponds to the center of the darkest region with $r_0 = 2.199$ in (a) and $r_0=2.592$ in (b).  The interval of contours is $8.265\times 10^{-4}$ in (a) and is $6.472\times 10^{-4}$ in (b).
}
\end{figure}
We recall that a Laughlin-type state with $\nu=1/p$ belongs simultaneously to the series $S_N$ and $\tilde{S}_N$.
Then $n(z,z')$ should be a superposition of pentagon- and hexagon-like patterns. 
To demonstrate this, the pair correlation function for $p=3 \ (M=45)$ is calculated and is shown in Fig.5 . It clearly confirms the expectation.
\begin{figure}
\figureheight{2cm}
\caption{Contour map of the pair correlation function $n(z,z')$ for $M=45$ with $z'=2.915 i$.  The interval of contours is $5.045\times 10^{-4}$.}
\end{figure}

Similar calculation is also performed for $N=7$.  Figure 6 shows the result for $M=39$ which belongs to the series $\tilde{S}_N$.  It shows the hexagon-type pattern as expected.  On the contrary, we have confirmed that non-magic-number states such as $M=35 \ (N=7)$ and $M=34 \ (N=6)$ do not show a regular polygonal pattern.
\begin{figure}
\figureheight{2cm}
\caption{Contour map of the pair correlation function $n(z,z')$ for $M=39, N=7$ with $z'=2.442 i$.  The interval of contours is $1.343\times 10^{-3}$.}
\end{figure}

\section{Interpretation of Magic Numbers}

\subsection{Geometrical interpretation}

As we have seen in the previous section, all the magic number states with $N=6$ belong either to the series $S_6$ or $\tilde{S}_6$.
In this section, we discuss the relation between the total angular momentum and the symmetry of the many-electron wave function.  The argument extends the previous one which explains the series $S_N$.\cite{youyan}
Let us consider the wave function 
$\Phi(z_{1},z_{2},\cdots,z_{N})$
in the case where the coordinates correspond to apexes of a regular polygon: $z_n = R\exp(i n\phi)$ with $n= 1,2,\cdots,N$ and $\phi=2\pi/N$. 
By applying $\exp(i\hat{M}\phi)$ where $\hat{M}$ is the total angular momentum operator, we obtain 
\begin{eqnarray}
\begin{array}{ll}
\lefteqn{ }\exp(i\hat{M}\phi)\Phi(z_{1},z_{2},\cdots,z_{N}) =\Phi(z_{N},z_{1},\cdots,z_{N-1}) \nonumber \\
=(-1)^{N-1}\Phi(z_{1},z_{2},\cdots,z_{N}).
\end{array}
\end{eqnarray}
The second equality follows from the antisymmetry of the fermion wave function.
Then we get
\begin{equation}
[\exp(iM\phi)+(-1)^{N}]\Phi(z_{1},z_{2},\cdots,z_{N})=0.
\end{equation}
In order for the regular polygonal pattern with $N$ apexes to be realized,
$\Phi(z_{1},z_{2},\cdots,z_{N})$ must be nonzero.
This gives a selection rule on $M$ as follows:
\begin{equation}
M= N(j+\frac{1}{2}), \quad (j=0,1,\cdots)
\end{equation}
if $N$ is even  and 
\begin{equation}
M=N j, \quad (j=0,1,\cdots)
\end{equation}
if $N$ is odd.  
The Pauli principle in addition requires $M \ge N(N-1)/2$.
This selection rule leads to the series $S_N$.

In order to examine the case of the polygonal pattern with $N-1$ apexes for $N$ electrons, we put $z_N =0,\ z_n = R\exp(i n\phi ')$ with $n= 1,2,\cdots,N-1$ and $\phi '=2\pi/(N-1)$.
Then we obtain
\begin{equation}
[\exp(i\frac{2\pi M}{N-1})+(-1)^{N-1}]\Phi(z_{1},z_{2},\cdots,z_{N-1},0)=0.
\end{equation}
This leads to another selection rule:
\begin{equation}
 M=(N-1) j, \quad (j=0,1,\cdots)
\end{equation}
if $N$ is even, and 
 \begin{equation}
 M=(N-1) (j+\frac{1}{2}), \quad (j=0,1,\cdots)
 \end{equation}
if $N$ is odd. 
This selection rule leads to the series $\tilde{S}_N$.

These rules nicely explain the occurrence of the magic number states in the case of $N\leq 6.$ 
However for larger $N$, the configuration with $N$-apex pattern costs more energy than another apex pattern with one or more electrons in the interior.  This is the reason why the magic number series $S_7$ is hardly seen in Fig.1(b).  
The exceptional appearance of $S_7$ is the magic number state $M=28$ with $N=7$.  The stability of the state, however, is better explained in terms of the CF picture as discussed below.
                                 
Now we turn attention to the $\nu=1/3$ Laughlin-type state which belongs to the series $S_N$ and $\tilde{S}_N$ simultaneously. 
Since this state has superposition of the polygonal pattern with $N$ apexes and the one with $N-1$ apexes, an extra stability is expected due to their resonant energy.

\subsection{Composite fermion picture}

In ref.\citen{jain6} an interpretation of magic numbers in terms of the CF picture is presented.  In this picture the wave function is constructed in the following form:
\begin{equation}
\Psi _{CF} (z_1,\ldots ,z_N)= \prod _{i<j} (z_i-z_j)^{2q} {\cal P} \Psi_0  (z_1,\ldots ,z_N),
\label{CF}
\end{equation}
where $\Psi_0  (z_1,\ldots ,z_N)$ is a wave function of free electrons, and $q$ is a natural number.  The Jastrow-type factor in eq.(\ref{CF}) is interpreted as binding a magnetic flux with strength $2q$ to each electron, hence the name of the CF.\cite{jain4}
Although $\Psi_0  (z_1,\ldots ,z_N)$ is not restricted to the LLL, the projection operator ${\cal P}$ picks out only such component that belongs to the LLL.
Thus the total angular momentum $M$ of $\Psi _{CF} (z_1,\ldots ,z_N)$ is the sum of the part $M_0$ associated with  $\Psi_0  (z_1,\ldots ,z_N)$ and that coming from the Jastrow-type factor.  Namely we obtain 
\begin{equation}
M=qN(N-1) +M_0.
\end{equation}

According to ref.\citen{jain6} the magic number corresponds to such $\Psi_0  (z_1,\ldots ,z_N)$ that has compact occupation of each Landau level from the lowest possible angular momentum.  
In the case of $N \leq 5$ the CF picture explains all the magic numbers.\cite{jain6}  The CF magic numbers for $N=5$ are the almost the same as the series $S_5$ with additional ones $M=18$ and 22 from the series $\tilde{S}_5$. 
For $N=4$ all magic numbers in $S_4$ are also given by the CF picture, while the CF picture gives $M=12$ as the only magic number in the series $\tilde{S}_4$.
The series $S_N$ with the electron number $N \leq 5$ have been studied by other authors \cite{trugman,maksym,maksym2} as well.  

For the case of $N=6$ and $q=1$, the state with $M=45$ has six electrons in the LLL.  This state has $M_0 =15$.  The other extreme case is the one where each of the six lowest Landau levels has only one electron, which we write as $(1,1,1,1,1,1)$, resulting in $M_0=-15$.  If one excludes an occupation where a higher LL has more electrons than a lower one, the number of compact states is counted as 11 for $N=6$.  However, both occupations $(3,1,1,1)$ and $(2,2,2)$ give $M=27$, and both $(4,1,1)$ and $(3,3)$ give $M=33$.
The set of the magic numbers with $q=1$ are then 
$$
M = 15, 21, 25, 27, 30, 33, 35, 39, 45.
$$
All these states belong to either $S_6$ or $\tilde{S}_6$.  
With $q=2$ the lowest magic number is $M=45$ and the second lowest is 51. A nice feature of the CF picture is to reject $M=20$ as a magic number in accordance with the numerical result in Fig.1(a).  
On the other hand, the numerical result shows clear cusps at $M=40$ and 50 in contradiction to the CF picture.

Similar analysis is carried out for $N=7$.  The number of compact states is 15 for each $q$ and the magic numbers with $q=1$ are given by
$$
M = 21, 28, 33, 35, 36, 39, 41, 42, 43, 45, 48, 49, 51, 56, 63.
$$
By comparing with Fig.2(b) we see that many of them correspond to cusps in the ground state energy.
However, the numerical result shows clear cusps also at $M=57$ and 69 both of which do not correspond to compact states.
As $N$ increases, the magic numbers predicted in the CF picture appear more densely than given in the series $S_N$ and $\tilde{S}_N$.  
Some of them appear clearly in the numerical result, while others have almost no cusp.  Unfortunately the relative stability among magic number states is not given in the CF picture.
We remark that among the magic numbers which are not explained in the CF picture, $M=40$ for $N=6$ and $M=57$ for $N=7$ are both next to smaller magic numbers, 39 and 56.


\section{Discussion and Conclusion}

We have seen that there are apparently conflicting interpretations of magic number states: the geometrical one which emphasizes the real space configuration, and the CF one which emphasizes the compact occupation of the one-body angular momentum.  
In the case of small $M$, the CF picture works better in rejecting $M=20$ with $N=6$, or $M=27$ with $N=7$.  However, in the case of large $M$, the geometrical interpretation works better in explaining $M=40$ and 50 with $N=6$, and $M=57$ and 69 with $N=7$. 
In the intermediate values of $M$, many of the magic numbers are common to both the geometrical and the CF pictures.   By combining both pictures we can account for all the magic numbers. 

It should be noted that the smaller $M$ corresponds to higher density of electrons in the macroscopic limit.   Thus the crossover in the effectiveness of the geometrical and the CF interpretations seems to reflect the transition from a FQH liquid at high density to the Wigner solid at low density in the macroscopic limit. 
The success of both interpretations in the intermediate range of $M$ is intriguing.  This suggests a smooth crossover for finite $N$ from the incipient FQH liquid to the incipient Wigner solid with decreasing density.

In this connection we remind the early work \cite{kivelson} which argues that particular densities for the stable FQH states also stabilize the Wigner crystal with large zero-point motion and the ring exchange.  It is thus not surprising that some of the electronic states in the quantum dot connect both to the FQH state and the Wigner crystal.
Experimentally, a particular advantage of the quantum dot is the wide range of controllable electron numbers $N$.  Thus one may observe how the electronic states change from the atomic type to the macroscopic type as $N$ is increased.  This feasibility is in strong contrast to genuine atoms.

In summary, we have investigated the quantum dot system in a strong magnetic field by numerical diagonalization of the Hamiltonian.  We have provided unified understanding of all the magic numbers in the total angular momentum.

\section*{Acknowledgment}

The authors are grateful to Y. Kato, T. Nihonyanagi, S. Tokizaki and H. Yokoyama for fruitful discussions.
The numerical calculations are performed by SX3-44R at the computing center of Tohoku University.
This research was supported by a Grant-in-Aid for Scientific Research on Priority Area from the Ministry of Education, Science, Sports and Culture.

\end{document}